\documentclass{emulateapj}

\usepackage{graphicx}
\usepackage{apjfonts}

\newcommand{\vl}{$v_{\rm LoS}$}
\newcommand{\ms}{${\rm m}\, {\rm s}^{-1}$}
\newcommand{\kms}{${\rm km}\, {\rm s}^{-1}$}
\newcommand{\degree}{^{\circ}}

\newcommand{\myemail}{jti@iaa.es}

\shorttitle{Spectropolarimetry with visible lines}
\shortauthors{Del Toro Iniesta, Orozco Su\'arez, \& Bellot Rubio}

\begin{document}

\title{On spectropolarimetric measurements with visible lines}

\author{J.C. del Toro Iniesta}
\affil{Instituto de Astrof\'{\i}sica de Andaluc\'{\i}a (CSIC), Apdo. de Correos 3004, E-18080 Granada, Spain}
\email{\myemail}

\author{D. Orozco Su\'arez}
\affil{National Astronomical Observatory of Japan, Mitaka, Tokyo 181-8588, Japan}
\email{d.orozco@nao.ac.jp}

\and

\author{L.R. Bellot Rubio}
\affil{Instituto de Astrof\'{\i}sica de Andaluc\'{\i}a (CSIC), Apdo. de Correos 3004, E-18080 Granada, Spain}
\email{lbellot@iaa.es}

\begin{abstract}
The ability of new instruments for providing accurate inferences of vector magnetic fields and line-of-sight velocities of the solar plasma depends a great deal on the sensitivity to these physical quantities of the spectral lines chosen to be measured. Recently, doubts have been raised about visible Stokes profiles to provide a clear distinction between weak fields and strong ones filling a small fraction of the observed area. The goal of this paper is to give qualitative and quantitative arguments that help in settling the debate since several instruments that employ visible lines are either operating or planned for the near future. The sensitivity of the Stokes profiles is calculated through the response functions (e.g. Ruiz Cobo \& Del Toro Iniesta, 1994). Both theoretical and empirical evidences are gathered in favor of the reliability of visible Stokes profiles. The response functions are used as well for estimating the uncertainties in the physical quantities due to noise in the observations. A useful formula has been derived that takes into account the measurement technique (number of polarization measurements, polarimetric efficiencies, number of wavelength samples), the model assumptions (number of free parameters, the filling factor), and the radiative transfer (response functions). We conclude that a scenario with a weak magnetic field can reasonably be distinguished with visible lines from another with a strong field but a similar Stokes $V$ amplitude, provided that the Milne-Eddington approximation is good enough to describe the solar atmosphere and that the polarization signal is at least three or four times larger than the typical rms noise of $\, 10^{-3} I_{\rm c}$ reached in the observations.
\end{abstract}


\section{Introduction}
\label{sec:intro}

A debate has been maintained for quite long about the capability of visible spectral lines to accurately measure weak magnetic fields and about whether can they be mistakenly confused with strong fields and low filling factors. The discussion often reaches levels for the whole community to rethink the current means for retrieving the magnetic properties of the solar photosphere. Should such confusions be as easy and widespread as suggested by several papers, the usage of visible spectropolarimeters and magnetographs would advisably be stopped. As a matter of fact, however, modern visible spectropolarimeters, e.g. the {\em Hinode} \citep{2007SoPh..243....3K,2008SoPh..249..167T} spectropolarimeter \citep{2001ASPC..236...33L}, are providing many new important discoveries whose reliability seems not to be easily put in doubt. Here we try to show that this is indeed the case and that reasonable accuracy can be achieved with visible lines whereby strong and weak fields cannot be muddled up easily. 

Advances in infrared spectropolarimetric instrumentation have improved significantly our ability to infer the magnetic and dynamic properties of the solar atmosphere. The large wavelengths and Land\'e factors of some spectral lines make them excellent diagnostic tools because, for instance, some may display completely split $\sigma$ components for magnetic fields stronger than, say, 400 G. The advent of infrared data has brought about a controversy on the results obtained for internetwork fields when compared to those coming from visible-line observations. The visible Fe {\sc i} lines at 630 nm have been indicating a predominance of kG fields \citep{2000ApJ...532.1215S,2003ApJ...582L..55D,2004ApJ...616..587S}, whilst the infrared lines at 1565 nm suggested hG fields \citep{1995ApJ...446..421L,1999ApJ...514..448L,collados2001,2003A&A...408.1115K,2006ApJ...646.1421D}. Whereas a decade ago \cite{1998ApJ...494..453W} provided evidence on the accuracy of Milne-Eddington (ME) inversions of the pair of Fe  {\sc i} lines at 630 nm for weak fields, some recent works even raise serious doubts about the reliability of visible-line inferences through inversion \citep{2006A&A...456.1159M}. The suggestion by \citet{2003ApJ...593..581S} about a bias of visible lines to strong fields and infrared lines to weak fields is too disappointing because it somehow admits an inability of visible lines to diagnose an important fraction of the solar magnetism. There certainly are differences in the sensitivities of lines in both wavelength regions but this should not mean that we must abandon spectropolarimetric measurements in the visible. There is much more information available if simultaneous observations in the visible and the infrared are used \citep[e.g.,][]{2006ApJ...649L..41C,2007A&A...475.1067C,2008A&A...477..273C, 2008A&A...477..953M,2008A&A...480..825B}. But is there any hope that visible line observations can provide reliable inferences when weak fields are observed? A solution to this controversial dispute is urgently needed because infrared wavelengths are not available to many polarimeters. Indeed some new instruments have been devised to use visible lines, namely, the Helioseismic Magnetic Imager \citep{2002AAS...200.5604S} aboard {\em Solar Dynamics Observatory}, the Imaging Magnetograph eXperiment \citep{2004SPIE.5487.1152M} aboard {\em Sunrise} \citep{2008AdSpR..42...70T}, and the Polarimetric and Helioseismic Imager aboard {\em Solar Orbiter}.  The results by \citet{2007ApJ...662L..31O,2007ApJ...670L..61O,2007PASJ...59S.837O} from a thorough analysis of magnetohydrodynamic simulations and real data have shed some light to the problem because their results indeed reconcile visible and infrared measurements. They concluded on the reliability of visible-line results if enough spatial resolution is available and stray light from the surroundings is appropriately taken into account.

The present paper tries to provide additional arguments to help discriminate whether or not visible spectropolarimetry is able to disentangle the two extreme cases when inferences are made through inversion techniques of the radiative transfer equation.  In Sect. \ \ref{sec:modelscenario} we clarify a number of points regarding inversion techniques and the underlying model atmosphere and physical assumptions; in Sect.\ \ref{sec:minimum} we discuss on the intrinsic ability of visible lines to distinguish weak from strong fields; in Sect.\ \ref{sec:uncertainties} we calculate the uncertainties of regular ME inversions if only noise is a limiting factor. These uncertainties can be considered as detection thresholds for the corresponding model parameters. Finally, in Sect.\ \ref{sec:conclu} we summarize the conclusions of this work.

\section{Models, scenarios, and inversions}
\label{sec:modelscenario}

The vocabulary related to inversion techniques, and to astrophysical measurements in general, is sometimes misleading to some extent. We often speak about model atmospheres, about scenarios, about hypotheses without clarifying their meanings. Let us, therefore, specify what we mean in this paper for each term and how we use all the concepts involved. On the other hand, practice teaches that failures occur from time to time and inversions might eventually not be satisfactory. Why does an inversion code occasionally fail to find a reliable solution? Is there anything intrinsic to the procedure that disable the success capabilities? We address these two issues in this section

First, a distinction should be made between a model atmosphere, that is, the set of parameters that fully characterize the solar medium from which the Stokes spectrum is coming, and the physical scenario, i.e., the set of physical assumptions about the composition of such model atmosphere and hence the means for defining which model parameters are relevant. For example, a physical scenario can be such that the physical quantities are assumed to be constant with depth; if we include the effects of stray light or of a partial filling of the resolution element, the corresponding model atmosphere can be the set of nine ME parameters\footnote{$S_0$ and $S_{1}$ for the constant term and the slope of the source function, $\eta_{0}$ for the line-to-continuum absorption coefficient ratio, $\Delta\lambda_{\rm D}$ for the Doppler width of the line, $a$ for the damping parameter of the line, $B$, $\gamma$, and $\varphi$ for the strength, inclination, and azimuth of the vector magnetic field, $v_{LoS}$ for the line-of-sight velocity, and $\alpha$ for the magnetic filling factor.} plus the filling fraction of the magnetic component. A given set of Stokes profiles could be inverted after assuming two different scenarios; for instance, a second scenario would be that the magnetic and velocity field vary with depth through the atmosphere. Should both of them produce equally good fits to the observations (similar values of the merit function), the Occam's razor principle is usually applied: the simpler, the better. In other words, practice recommends to be careful with involved scenarios (e.g. gradients along the line of sight or several components within the resolution element) whenever a simple one (e.g. ME) produces good fits to the observations. Of course, fit quality is a subjective concept but can be quantified in terms of ($\chi^{2}$) merit-function values: we can set a threshold for the differences between observed and synthetic (best-fit) profiles; if  $\chi^{2}$ reaches smaller values than this threshold, we can say we are satisfied with the inversion. Let us explain our point otherwise: imagine that the second scenario does not lead to a clear minimum of the merit function because two different sets of model parameters yield no significantly different $\chi^{2}$. In such a case, we cannot say that the observables (e.g., the visible line Stokes profiles) do not provide reliable diagnostics unless we try with simpler scenarios like the first one with constant physical quantities. It may well be that this scenario provides good fits to the data.\footnote{By good fits we understand both $\chi^{2}$ values below the established threshold and solution uniqueness or independence of the initial guess.} In Sect. \ref{sec:realcase} we show one such case.

Second, there are only two reasons for a Levenberg-Marquardt inversion code to fail, namely, that the merit function has no well-defined global minimum in the hyperspace of parameters or that the updating strategy for the model atmosphere\footnote{We hereafter restrict the discussion to Levenberg-Marquardt algorithms as are among the most used ones in practice. The fudge parameter $\lambda$ is the technical means for these algorithms to update the model atmosphere by appropriately combining the advantages of the inverse-Hessian and the steepest-descent methods. The latter is in general better when far from the global minimum and the second when close to it.} is not well suited to the problem. Let us discuss both reasons separately.

Consider two or more model atmospheres within the same scenario. If they produce equally good fits (i.e., equally low values of the $\chi^2$ merit function), then either the observational noise is such that hides the true minimum, or the underlying physical scenario is not appropriate to describe the solar atmosphere, or both. We take for granted that noise hampers the measurements. Indeed it governs the threshold above which the model parameters can be reliably inferred (see Sect. \ref{sec:uncertainties}).  Discarding a situation where the noise is too high, however, let us speak about the scenario. By ambiguity of the scenario we mean that the physical constraints behind it can be such that the number of free parameters is too high for the information content available from the observables or that some parameters cannot be reliably retrieved. For instance, \citet*{2006A&A...456.1159M} have shown that the pair of Fe {\sc i} lines at 630 nm is not able to provide a single model for a scenario in which two depth-dependent atmospheres, one magnetic and another non magnetic, fill a spatial resolution element of about 1\arcsec. This can be an example of the first case: several combinations of the atmospheric quantities yield the same $\chi^2$ value. As for the second case, a good example is found in the well-known trade-offs among $\eta_0$, $\Delta \lambda_D$, and $a$ of ME inversions \citep[see][Orozco Su\'arez \& Del Toro Iniesta 2007]{1998ApJ...494..453W}: several sets of such three parameters may give fits with the same quality without changing the magnetic and velocity parameters. This means that the thermodynamic parameters of a ME atmosphere are not reliable while the magnetic and dynamic are.

Let us suppose now that the physical scenario explains the observations, that is, that a clear global minimum should exist. To reach it, the usual strategy is to increase or decrease the $\lambda$ parameter by a factor 10, depending on the divergence or convergence of the algorithm. If the code reaches a local minimum, then the fudge parameter has to be large enough to favor bigger perturbation steps and overcome that minimum. Sometimes, however, the updating strategy is such that the fudge parameter is unable to get rid of the local minimum. For example, in the MILOS program \citep{2007A&A...462.1137O} $\lambda$ monotonically decrease until it starts oscillating and gets trapped. Alternative updating strategies as well as initial values for the fudge parameter can be devised to increase the efficiency of convergence, but ultimately they are somehow based on the user's experience. The only automatic option is to stop the inversion procedure as soon as oscillations appear.

No major technical problem seems to exist, then, provided that an appropriate scenario and a good fudge-parameter strategy have been selected for inversion codes to reach good fits to the observations. The distinction between different model atmospheres within a single physical scenario (e.g. the distinction between weak and strong magnetic fields) only depends on the information content in the Stokes profiles. Such a content depends in its turn on the spatial resolution (filling fraction of pixel occupation) and on the noise of the observations. The conclusions, hence, cannot be general for given spectral lines but specific for a given observation. For instance, \citet{2003A&A...406..357B} distinguish between weak and strong fields provided the observational differences are significantly above the noise. 

\section{Detectable signatures of weak and strong fields}
\label{sec:minimum}

This section is devoted to show the differences between Stokes profiles formed in atmospheres with weak magnetic fields plus large filling factors and strong magnetic fields but small filling factors. Such differences enable the inversion codes to reliably distinguish between the two cases as we are going to see through a double approach. First, we use synthetic profiles and theoretical arguments. Second, a sample case with actually observed Stokes profiles is shown. They have been inverted with the MILOS code. This code deals with individual lines but, like other ME inversion codes existing in the literature \citep[e.g.][]{1987ApJ...322..473S}, MILOS can invert simultaneously specific pairs of lines. This is the case of the two Fe {\sc i} lines at 630 nm. On the one hand, inversion of lines not being normal Zeeman triplets like that at 630.15 nm is possible by computing the full Zeeman pattern and considering the components altogether, following \citet{1977SoPh...55...47A}. On the other hand, the two lines belong to the same multiplet and, hence, are fairly similar in everything except for their line strength. This property allows the user not to increase the number of free thermodynamic parameters by simply considering the ratio of their oscillator strengths \citep{1987ApJ...322..473S} and, thus, of their $\eta_{0}$'s: just one $\eta_{0}$, one $a$, one $\Delta\lambda_{\rm D}$, one $S_{0}$, and one $S_{1}$ are retrieved consistently with one $B$, one $\gamma$, one $\varphi$, and one $\alpha$. In any other technical aspect, MILOS is a regular Levenberg-Marquardt non-linear, least-squares algorithm that inverts the full radiative transfer equation under the assumption of the ME approximation using measurements of the four Stokes parameters to better constrain the inversion.\footnote{It is programmed in IDL language and is publicly available on our Website, http://spg.iaa.es/download.asp.}

\subsection{The synthetic case}
\label{sec:syntheticcase}

\begin{figure}[!ht]
	 \centering
\resizebox{\hsize}{!}{\includegraphics[width=\textwidth]{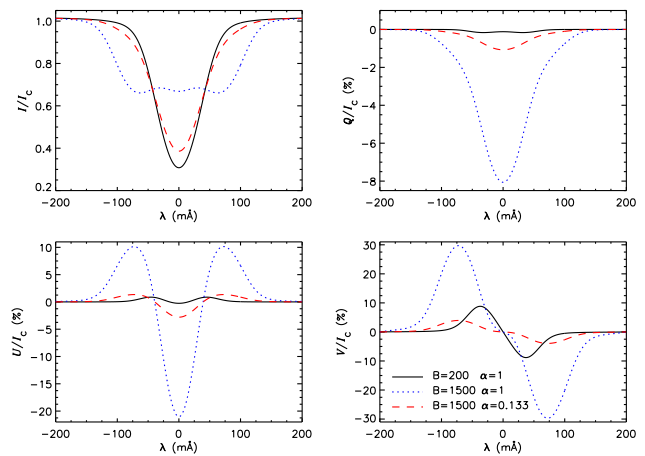}}
  \caption{Emergent Stokes profiles from three different ME atmospheres
  with the same thermodynamic parameters. Differences lie on their magnetic field strengths and filling factors. Check line coding in the figure inset (right, bottom panel).}
  \label{fig:stoprofiles}
\end{figure}

Consider two ME atmospheres whose thermodynamic parameters are the same. Let us take for instance those coming from a fit to the intensity profile of the Fe\ {\sc i} line at 630.25 nm in the Fourier Transform Spectrometer Atlas of the quiet Sun \citep{Brault1,Brault2}. Two different atmospheres will be built by assuming a different magnetic field strength: 200 and 1500 G. The magnetic inclination and azimuth, $\gamma$ and $\phi$, are taken equal to 45\degr for both model atmospheres. The specific values have nothing to do with the qualitative results and, in fact, intend to be representative of the general problem. From these two atmospheres different Stokes profiles emerge as can be seen in Fig.\ \ref{fig:stoprofiles}. Solid and dotted lines correspond to the weak and strong field cases, respectively. The shapes of the profiles are clearly dissimilar and the Zeeman splitting is more conspicuous for the strong-field profiles. A filling factor unity is assumed in both cases. The differences are so significant that distinguishing between the two cases is fairly easy. Inversion techniques should have no problem in inferring either of the two models. 

It could be argued that a filling factor $\alpha = 0.133$ in the strong-field case would be equivalent to the weak-field atmosphere because the magnetic flux is the same in the two models. However, this would disregard the enormous absolute differences in the profile shapes: as can also be seen in Fig.\ \ref{fig:stoprofiles}, they can be as large as 1\% in Stokes $Q/I_{\rm c}$, as 2\% in Stokes $U/I_{\rm c}$, and as 8--10\% in Stokes $I/I_{\rm c}$ and $V/I_{\rm c}$. Hence, the differences are between one to two orders of magnitude larger than the typical noise ($10^{-3} I_{\rm c}$) achievable by modern spectropolarimeters and are thus discernible by current inversion codes. Of course, our conclusion depends on the validity of the scenario as the belief on a better accuracy for the flux determination than for the field strength determination roots in the proportionality between magnetic flux and Stokes $V$. This proportionality is true in a very narrow range of field strengths, however: we know, indeed, that strict proportionality only takes place while second-order effects are negligible; but the mere existence of linear polarization breaks this condition: Stokes $Q$ and $U$ are of second order \citep{1992soti.book...71L}. Both circular and linear polarization scale with $\alpha$ in the weak field regime but while $V$ scales with $B$, $Q$ and $U$ go as $B^{2}$ and are even more sensitive to changes in the magnetic field strength. This means that wherever we measure linear polarization, the proportionality is not exact. Therefore, second-order effects are enough for getting accurate inferences provided they are bigger than the noise. This is the case, for example, of the weak fields detected in the internetwork by the spectropolarimeter aboard the {\em Hinode} satellite \citep{2007PASJ...59S.571L, 2008ApJ...672.1237L, 2007ApJ...670L..61O}, that had hitherto gone unnoticed from the ground. Even in the case of strictly longitudinal fields (i.e. with no linear polarization), proportionality breaks down soon. Stokes $V$ ceases to scale with field strength at fairly moderate strengths: a linear and a parabolic fit to the Stokes $V$ amplitude as a function of the (longitudinal) field strength differ by 1-2\% of the continuum intensity between 0 and 400 G. For 500 G, the differences reach a 6\%.\footnote{The quoted differences correspond to the following ME parameters fixed:  $\eta_{0} = 22$, $\Delta{\lambda_{D}} = 30$ m\AA, $a = 0.03$, and $v_{LoS} = \gamma = \varphi = 0$. The values are typical for visible lines like those of Fe {\sc i} at 630 nm. Small changes in these parameters give very similar results.} All these differences are significantly larger than the typical noise of $10^{-3}\: I_{c}$. Hence, as long as we have several wavelength samples (i.e. unless we perform single-point magnetometry), the chances of distinguishing non linearities are quite high. Numerical tests of the ability of ME inversions of the Fe {\sc i} lines at 630 nm have already been carried out by \citet{1998ApJ...494..453W} with synthetic profiles in both constant and depth-dependent atmospheres and by \citet{2007ApJ...662L..31O} with synthetic profiles from numerical MHD simulations of the photosphere. We further check and confirm the reliability of such inversions with the Monte-Carlo simulations we report in the Appendix. There, a total of 10$^5$ ME profiles have been inverted after adding them random noise with a rms value of $10^{-3}\: I_{c}$.

\begin{figure}
  \centering
  \plotone{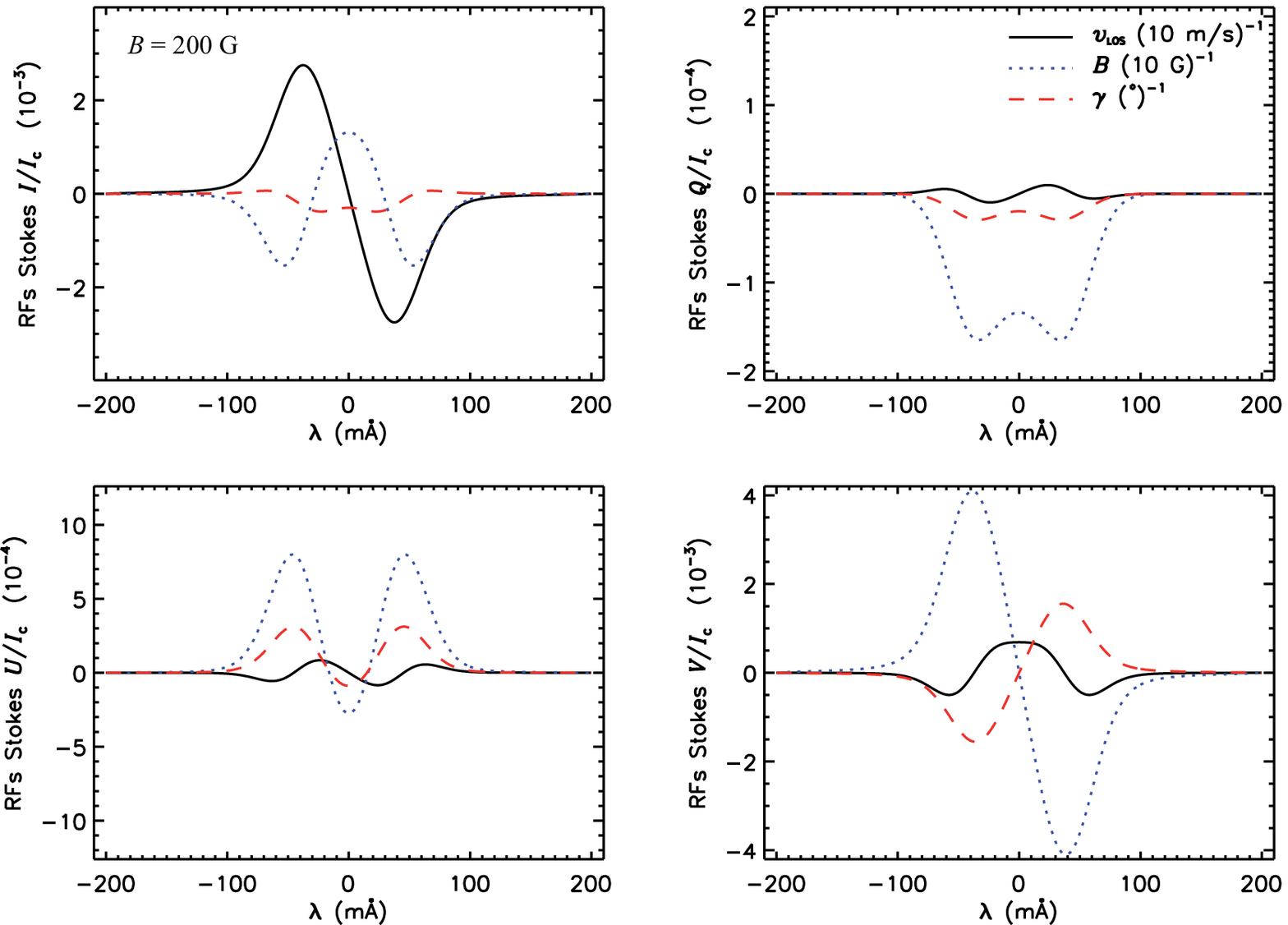}
  \plotone{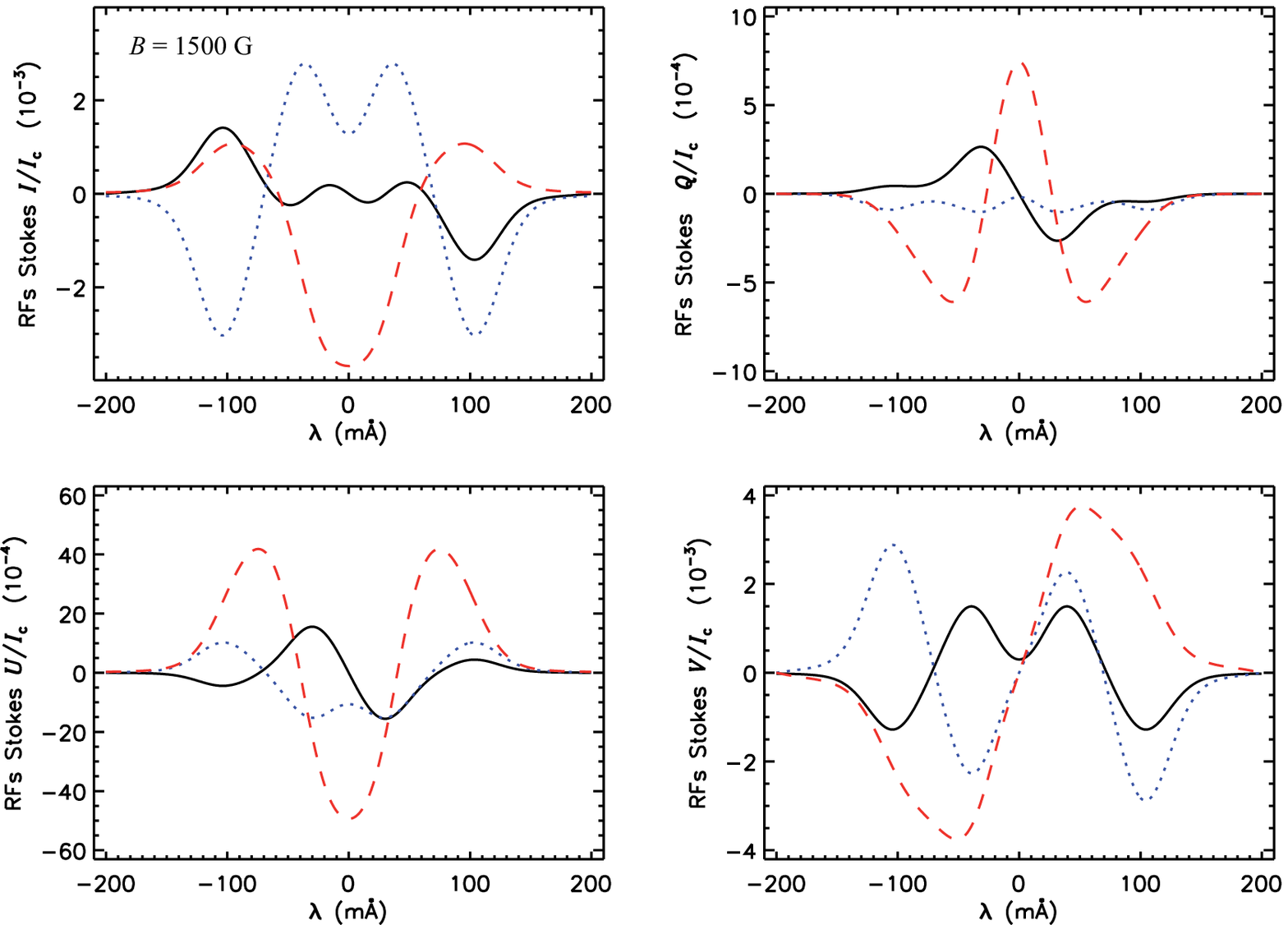}
  \caption{Response functions of the Fe {\sc i} Stokes profiles at 630.25 nm (normalized to the continuum intensity) for the weak- (four top panels) and strong-field (four bottom panels) ME models versus wavelength. RFs to $v_{\rm LoS}$ perturbations (solid, black lines), to $B$ perturbations (blue, dotted lines), and to $\gamma$ (red, dashed lines) are presented. Perturbations of 10 \ms, of 10 G, and of 1\degr are assumed.}
  \label{fig:Fig_B200-B1500}
\end{figure}

The most important feature from the measurement point of view, however, is the sensitivities of the profiles as given by response functions \citep{1994A&A...283..129R}. Response functions (RFs) of the Fe {\sc i} line at 630.25 nm to perturbations of the LoS velocity (solid lines), of the magnetic field strength (dotted lines), and of the magnetic inclination (dashed lines) are plotted versus wavelength in Fig.\ \ref{fig:Fig_B200-B1500}. The RFs for the weak-field model are in the four top panels and for the strong-field model in the four bottom panels. Note that the scales are the same for each Stokes parameter in the two models. Units have been selected to represent perturbations of 10 \ms, of 10 G, and of 1\degr, respectively, since the response to a 1\degr perturbation can be as large as (or even larger than) those to 1 G and 1\ms. The spectral line responds differently in the two models to the same perturbation in the atmospheric parameters. Moreover, the various wavelengths react differently to the same perturbation in the two atmospheres: both the shape and the amplitudes of the RFs are different. This is the key that allows inversion codes to discriminate between the two models. 

The relative accuracies of the inferences in the two cases are determined by their relative sensitivity or RF values. The difference in maximum sensitivities, however, is small: as seen in Fig.\ \ref{fig:Fig_B200-B1500}, 3 G can be as significant in one of the models as 1 G is in the other. The RF amplitude is even larger for Stokes $Q$ and $V$ in the weak model than in the strong model. Moreover, it is not only the maximum sensitivity what matters but the effect of the whole profile. Therefore, although the common belief is that strong fields are better retrieved than weak ones, inferences in the weak-field case can eventually be more accurate if the fields are not very horizontal as we are going to see in the next section. The point we want to stress in here is that the neat shape differences of the RFs help inversion codes disentangle strong from weak magnetic fields.\footnote{Inversion codes like SIR \citep{1992ApJ...398..375R} or MILOS are based on response functions. Other codes not explicitly programmed making use of RFs do indeed use them implicitly \citep[e.g.,][]{2003isp..book.....D}} This is a similar situation to that giving rise to a discrimination between the magnetic and dynamic parameters from the thermodynamic parameters in ME inversions \citep{2007A&A...462.1137O}.

\begin{figure}[!ht]
 \centering
\resizebox{1\hsize}{!}{\includegraphics[width=\textwidth]{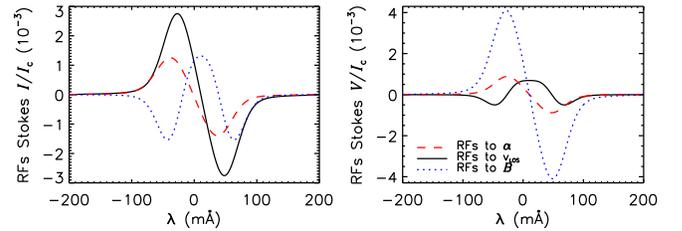}}
 \caption{Stokes $I$ (left) and $V$ (right) RFs to \vl (solid, black lines), to $B$ (dotted, blue lines), and to $\alpha$ (dashed, red lines) in the weak-field case. Perturbations of 10 \ms for \vl, of 10 G for $B$, and of 0.1 for $\alpha$ have been assumed.}
  \label{fig:rfalfa}
\end{figure}

The fact that filling factor and magnetic field strength influence differently the Stokes profiles can also be assessed with RFs. As shown in Fig.\ \ref{fig:rfalfa}, in the weak field model the sensitivity of Stokes $V$ to $\alpha$ perturbations has exactly the same shape than the sensitivity to $B$ perturbations (the same occurs for Stokes $Q$ and $U$). This is so because, at these strengths, $V$ scales with both $\alpha$ and $B$. (Both RFs have a $V$-shaped profile.) This is not the case, however, for Stokes $I$: the RFs of Stokes $I$ to $\alpha$ and $B$ are very different. In fact, while the response to $\alpha$ perturbations keeps a $V$-like shape, the RF to $B$ perturbations has three lobes.\footnote{Since the observed spectrum is a linear combination of the magnetic and non-magnetic atmospheres, the RF to $\alpha$ perturbations is simply the difference between the magnetic and non-magnetic Stokes profiles, as shown by \citet{2007A&A...462.1137O}.} Even in the weak field case, then, thanks to the Stokes $I$ behavior, the effects of $\alpha$ and $B$ can be discerned. Therefore, the reliability of the separation between the two cases roots in Stokes $I$, as first pointed out by \citet{2007PASJ...59S.837O}. The intensity cannot be fitted by modifying one of the two parameters regardless of the other.

\begin{figure}[!ht]
  \centering
\resizebox{\hsize}{!}{\includegraphics[width=\textwidth]{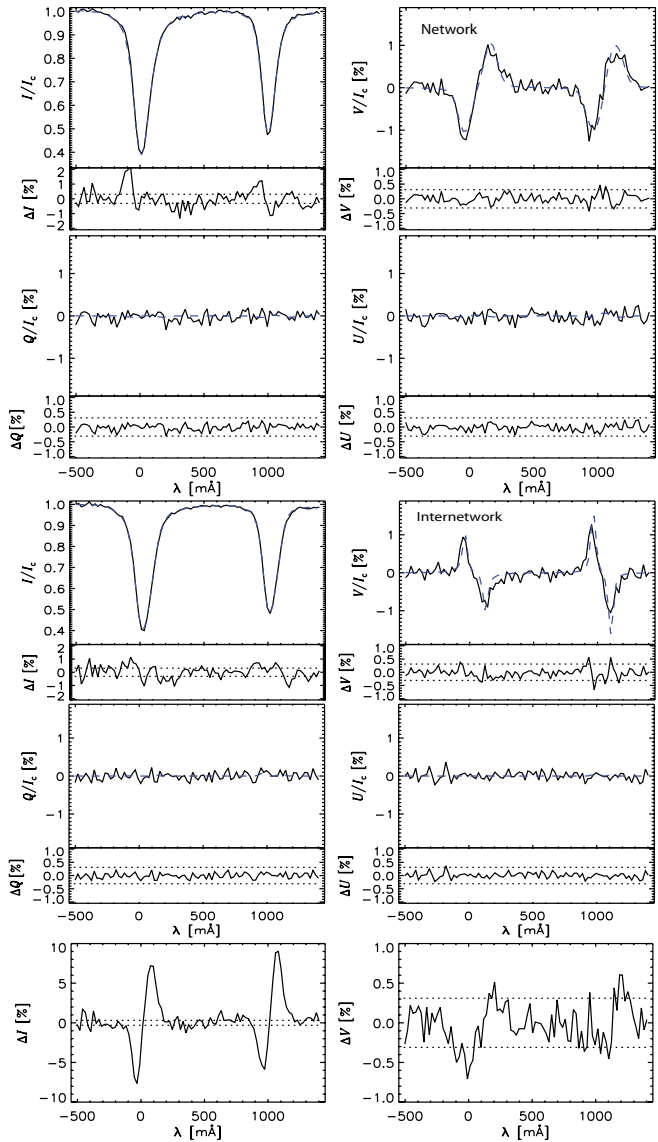}}
  \caption{Stokes $I$ and $V$ profiles of the Fe {\sc i} lines at 630 nm as observed with the spectropolarimeter aboard {\em Hinode} in a network (upper panels) and an internetwork (middle panels) points. Observations (solid lines) and fits (dashed lines) are displayed. Their differences are also shown at the bottom of each panel. The bottom panels show the difference (network $-$ internetwork) between the profiles of the two points. Note that the internetwork polarity has been artificially changed for calculating this difference.}
  \label{fig:realprofiles}
\end{figure}

\subsection{Observed Stokes profiles: the real case}
\label{sec:realcase}

Problems of distinction between the two cases invariably occur in real observations when polarization signals are weak. As an illustration, we have chosen the Fe {\sc i} lines at 630 nm as observed in two points, one from the network and another from the internetwork. The data come from the {\em Hinode} observations already used by \citet{2007PASJ...59S.571L,2008ApJ...672.1237L} and by \citet{2007PASJ...59S.837O,2007ApJ...670L..61O}. The circular polarization is less than or equal to 1\% of the continuum intensity in both points. The linear polarization is buried within the noise. The Stokes profiles of both points are shown in Fig.\ \ref{fig:realprofiles} (four first rows) along with their differences (bottom panels). The observed (solid line) profiles and the fits (dashed line) are included. The quality of the fits is very good as indicated by the residuals: those for the $V$ profiles hardly exceed the $3\sigma$ level marked by the horizontal dotted lines. ($\sigma = 10^{-3} I_{\rm c}$ is the typical noise level of the {\em Hinode} spectropolarimetric observations.) The ME inversions carried out with the MILOS program give $(160 \pm 2)\degree$ and $(50 \pm 2)\degree$ for the field inclination, $\gamma$; $0.130 \pm 0.005$ and $0.25 \pm 0.02$ for the magnetic filling factor, $\alpha$; $(2.8 \pm 0.1)$~\kms and $(2.0 \pm 0.1)$~\kms for the line-of-sight velocity, $v_{\rm LoS}$; and $(1400 \pm 200)$~G and $(280 \pm 80)$~G for the magnetic field strength, $B$. The uncertainties are obtained after a Montecarlo experiment repeating the inversion 1000 times and are indeed consistent with the analytic estimation of uncertainties that we describe in the following section. 

Although the amplitudes are fairly similar (their differences are larger than $3\sigma$ in just a few wavelength samples), regardless of the magnetic polarity, a net difference in the width of the Stokes $V$ lobes can easily be seen. Therefore, the two Stokes $V$ profiles are by no means proportional and the linearity hypothesis with the magnetic flux is not valid at all. In fact, the network profile corresponds to a flux four times larger than that of the internetwork profile. Nevertheless, although the subtle lack of proportionality should be detectable by the code, the conspicuous difference in Stokes $I$ (well above the $3\sigma$ level) points to very different field strengths. Since the field inclination angles are almost supplementary, the observed $I$ difference can only come from differences in $v_{\rm LoS}$, $\alpha$, $B$, and/or the thermodynamics. According to the inversion, the network point is shifted to the red with respect to the internetwork one. Hence, the expected asymmetric difference should be of {\em opposite} sign to that observed. The thermodynamic parameters are known to cross-talk among them \citep[e.g.][]{2007A&A...462.1137O} and, hence, their differences cannot give account of the profile differences. Therefore, most of them should come from $B$ and $\alpha$. In the previous section we have provided analytic evidence about neatly different sensitivities to $B$ and $\alpha$ perturbations but let us for the moment assume that this is not the case and that $B$ and $\alpha$ could have been fully confused: if $\alpha$ had to be 0.25 for both profiles, the network field strength would turn out to be 700 G, still very different to the internetwork value, and certainly not in the weak field regime. Finally, a further evidence is the Stokes $V$ amplitude ratio between the two lines of the Fe {\sc i} multiplet. Such a ratio suggests a strong field for the network profile and a weak field for the internetwork one.

\section{Uncertainties}
\label{sec:uncertainties}

As is well known, and already mentioned in Sect.\ \ref{sec:modelscenario}, noise can be a limiting factor to the accuracy of the results of any measurement technique and, in particular, of inversion codes. In this section we provide quantitative estimates of the uncertainties in the retrieved physical quantities induced by the noise if the observations are analyzed through a ME inversion code that assumes a magnetic atmosphere filling a fraction $\alpha$ of the resolution element. In such a case, the observed Stokes spectrum is classically expressed as
\begin{equation}
\label{eq:stokes}
\mathbf{I}_{\rm obs} = \alpha \mathbf{I_{\rm m}} + (1-\alpha)  \mathbf{I_{\rm nm}},
\end{equation}
where $\mathbf{I_{\rm m}}$ is the Stokes vector emerging from the magnetic atmosphere and $\mathbf{I_{\rm nm}}$ is obtained as a given average of the pixel surroundings.

Following \citet{1997ApJ...491..993S}, assume that all uncertainties in the $m$ physical parameters contribute in a similar amount to the final noise. (Actually, our assumption is that noise imparts equal, Gaussian-distributed uncertainties to the $m$ parameters that are sought.) In such a case, the variance of the $j$ wavelength sample in the $i$-th Stokes parameter\footnote{Index $i$ runs from 1 through 4, corresponding to Stokes $I$, $Q$, $U$, and $V$, respectively.} can be written as\footnote{Within the ME approximation, RFs are direct partial derivatives of the Stokes spectrum with respect to the model parameters \citep[see][]{2007A&A...462.1137O}. Obviously, only $\mathbf{I_{\rm m}}$ depends on the free parameters.}
\begin{equation}
\label{eq:uncertainties1}
\sigma_{i,j}^2 = m \alpha^2 \, (R_{i,j}^x)^2 \sigma_x^2,
\end{equation}
where $\alpha$ stands for the magnetic filling factor, $R_{i,j}^x$ is the RF of Stokes $i$ at wavelength $j$ to perturbations of the model parameter $x$, and $\sigma_x^2$ is the variance of that parameter. Summing up for all Stokes parameters and wavelengths, Eq.\ \ref{eq:uncertainties1} becomes
\begin{equation}
\label{eq:uncertainties2}
\sum_{i=1}^4 \sum_{j=1}^{n_\lambda} \sigma_{i,j}^2 = m \alpha^2
\sigma_x^2 \sum_{i=1}^4 \sum_{j=1}^{n_\lambda} (R_{i,j}^x)^2,
\end{equation}
where $n_\lambda$ is the number of wavelength samples.

According to \citet{2000ApOpt..39.1637D}, if all the modulated measurements ($n_p$) needed to derive the Stokes parameters have the same variance, $\sigma^2$, due, for instance, to photon noise, then 
\begin{equation}
\label{eq:variance}
\sigma_{i,j}^2 = \frac{1}{n_p} \frac{\sigma^2}{\epsilon_i^2}, \forall j=1, \ldots, n_\lambda,
\end{equation}
where $\epsilon_i$ is the polarimetric efficiency for the $i$-th Stokes parameter.

Using Eq.\ (\ref{eq:variance}), the uncertainty of the $x$ parameter can finally be cast as
\begin{equation}
\label{eq:variancex}
\sigma_x = \frac{\sqrt{n_\lambda \sum_{i=1}^{4} (1/\epsilon_i^2)} \, \sigma}{\alpha \sqrt{n_p m} \sqrt{\sum_{i=1}^4 \sum_{j=1}^{n_\lambda} (R_{i,j}^x)^2}}.
\end{equation}

\begin{figure}[!ht]
\hspace{-0.03\textwidth}
\resizebox{1.12\hsize}{!}{\includegraphics[width=\textwidth]{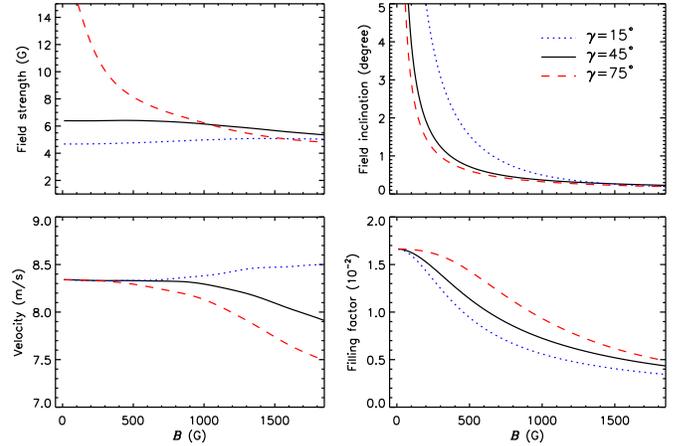}}
 \caption{Uncertainties for the retrieved field strength (top left panel), for the field inclination (top right panel), for the LoS velocity (bottom left panel), and for the filling factor (bottom right panel), according to Eq.\ (\ref{eq:variancex}), as functions of the magnetic field strength. Line colors and types indicate different field inclinations; see the inset.}
  \label{fig:Fig_B1500}
\end{figure}

The above formula gives an estimate for the noise-induced, i.e. random, effects and no systematic errors are included. The latter can be larger as already shown through numerical experiments by \citet{1998ApJ...494..453W}. Our formula, however, illustrates very well how the noise on the (modulated) polarization measurement influences directly the accuracy in any inferred parameter. Obviously, the better the polarimetric efficiencies of the instrument, the smaller the uncertainties. That is also the case for RFs: the larger the RFs (the sensitivity) the smaller $\sigma_x$. Values from Eq.\ (\ref{eq:variancex}) are plotted in Fig.~\ref{fig:Fig_B1500} for a range of magnetic field strengths (from 0 through 1850 G) and magnetic inclinations (15\degr\  in dotted lines, 45\degr\  in solid lines, and 75\degr\  in dashed lines). The model atmosphere is an average of those obtained after the inversion of more than two million quiet-Sun profiles \citep{2007ApJ...670L..61O}. $n_\lambda = 91$, $n_p = 4$, $\epsilon^2_i = 1, 1/3, 1/3, 1/3$, $\alpha = 1$, and $\sigma = 10^{-3} I_{\rm c}$. The best accuracies are obtained for vertical fields in all the strength regimes, except for the inclination that is (naturally) better determined when the field is close to the horizontal. The improvement on the velocity determinations for fields stronger than approximately 900 G for the $\gamma = 45\degr, 75\degr$ cases is due to a well-known shape effect \citep[e.g.][]{2005A&A...439..687C}: with these inclined magnetic fields, the $\pi$ component of the line starts to become prominent on Stokes $I$, making it deeper and narrower. That is also the reason for the overall increase in the \vl\  uncertainties with the field strength: the lines become broader as the strength grows. The neat decrease of $\sigma_B$ with $B$ in the  $\gamma = 45\degr, 75\degr$ cases is due to the appearance of Stokes $Q$ and $U$ signals. The slight but appreciable increase of the uncertainty with $B$ in the $\gamma = 15\degr$ case might be due to a net decrease in information: when the field is strong, larger strengths mean larger separation of the $V$ lobes but the variation in amplitude is small. Hence, the field strength uncertainties in the three inclination cases converge for strong fields. The behavior of the inclination and filling factors is supposed to be more natural: determinations are better when $B$ is large. Nevertheless, as we advanced in the previous section, it is the effect of all the four Stokes profiles that is relevant to the final inference; arguments based on just one Stokes parameter may fail. It is finally noteworthy that the accuracy in each parameter is inversely proportional to the magnetic filling factor, according to Eq.\ (\ref{eq:variancex}). Hence, the ordinate scale of the figure should be multiplied by 10 if $\alpha = 0.1$. In such a case, and even if our estimates were wrong by a 100\%, the expected uncertainties support our conclusion that strong and weak fields can be distinguished with visible lines provided that a ME atmosphere partially filling the resolution element can be assumed as a model of the solar photosphere that is observed with a noise of the order of $10^{-3} I_{\rm c}$.

\section{Conclusions}
\label{sec:conclu}

In this paper we have gathered evidence in favor of the ability of visible Stokes profiles to provide reliable information about the magnetic and dynamic properties of the solar photosphere and, in particular, to distinguish between strong and weak magnetic fields with similar Stokes $V$ amplitudes. Our results complement those by \citet{2007ApJ...662L..31O,2007ApJ...670L..61O,2007PASJ...59S.837O} and agree with them. Most of the problems found by other authors \citep[e.g.][]{2006A&A...456.1159M} can be ascribed directly to using scenarios that are too complex for the interpretation of the observations. The additional information provided by infrared lines or by the combined use of visible and infrared lines enables us to attack the problem by assuming complicated scenarios with several components within the resolution element whose physical quantities vary with depth and so on. These scenarios may not be appropriate for visible lines or for observations with limited wavelength sampling. Such observations should be analyzed in terms of less ambitious (hence simpler) scenarios where, probably, the number of atmospheric components is reduced to one plus a stray-light contribution and where physical quantities are constant with depth. 

We have shown that linearity between Stokes $V$ and $\alpha$ and $B$ cannot be any longer an excuse for people not trusting the results based on the inversion of visible lines. The reason is four-fold: linearity between $V$ and $\alpha B\cos \gamma$ takes place in a very narrow range of values and cannot be taken for granted; circular polarization scales with $B$ whilst linear polarization does it with $B^2$: linear polarization helps a lot for measuring $B$; the sensitivities of Stokes $I$ to $B$ and $\alpha$ are conspicuously different in any field regime and help to disentangle the effects of both parameters; and the uncertainties in the model parameters induced by the noise of modern spectropolarimeters set detection thresholds that are small enough. (These theoretical thresholds are fully compatible with the numerical results from Monte-Carlo simulations.) Occasional mistakes (e.g., disproportionally large field strengths with extremely small filling factors) obtained with existing inversion codes cannot then be ascribed to intrinsic difficulties based on radiative transfer theory but likely to either specific numerical and/or programming problems, or to ambitious analyses with too many free parameters.

\section*{Appendix}
\label{appendix}

To further check the reliability of our ME inversions of the Fe~{\sc i} pair of lines at 630 nm, in this appendix we perform the following numerical experiment: first, we generate a reference set of Stokes profiles; second, we add Gaussian noise with a rms value of $10^{-3}\, I_{{\rm c}}$; and, finally, we invert the Stokes profiles with the MILOS code. According to Eq.\ \ref{eq:stokes}, a magnetic atmosphere occupies a fraction $\alpha$ of the resolution element and a non-magnetized one occupies the rest. As for the magnetic component, 100\,000 ME model atmospheres with a uniform random distribution of vector magnetic fields (field strengths from 0 to 3000~G and inclinations and azimuths from 0 to 180$\degree$) have bee used. The remaining model parameters are those coming from the inversion of the FTS atlas data as in the main text (Sect.\ \ref{sec:syntheticcase}). The filling factor has been considered with a uniform random distribution as well, with values from $\alpha=0$ to $1$. The non-magnetized atmosphere is assumed to be known a priory. Same thermodynamic parameters as those for the magnetic component have been assumed. An extra broadening to the profile using a macroturbulent velocity of 1~km~s$^{-1}$ has been used which is consistent with a contribution mostly coming from the surroundings of each pixel.

All inversions have been carried out with the following initialization: $\lambda_0=1$ (fudge parameter), $S_0=0.02$, $S_1=1$, $\eta_0=4$, $B=800$~G, $\gamma=45\degree$, $\chi=45\degree$, $\Delta \lambda_\mathrm{D}=26$~m{\AA}, $v_\mathrm{LOS}=0$ km~s$^{-1}$, $a=0.15$, and $\alpha=0.6$. We have used a maximum of 30 iterations per pixel. As is usual in practice, only profiles displaying at least one of the polarization signals larger than $3\,10^{-3}\, I_{{\rm c}}$ are inverted: 5\% of the profiles have hence been discarded.

\begin{figure*}[htbp]
\begin{center}
\resizebox{\hsize}{!}{\includegraphics{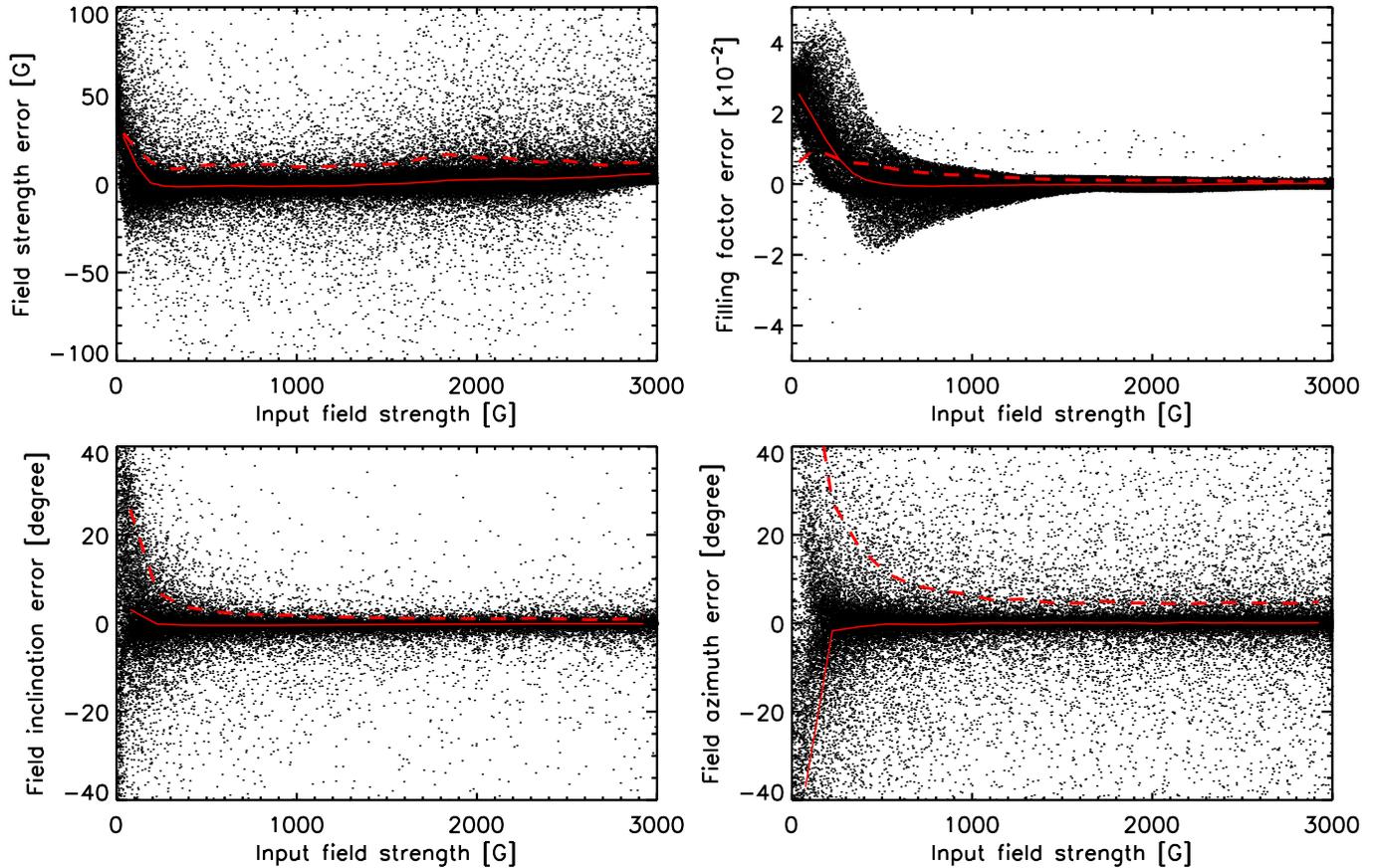}}
\caption{Differences between the inferred and the reference model parameters as a function of field strength. The solid and dashed lines indicate the mean and the rms values, respectively.}
\label{montecarlo}
\end{center}
\end{figure*}

The upper panels of Fig.~\ref{montecarlo} show the difference between the inferred parameters and the reference model. Each point corresponds to an individual trial of each of the 100\,000 ME inversions. Over-plotted are the corresponding mean (solid line) and rms values (dashed line). The panels show the values for the magnetic field strength, its inclination, azimuth, and the filling factor. The results indicate that the inferences are accurate enough for all the four model parameters. For fields of 100 and 200 G, the mean errors are 30 G and -3 G for the field strength, 1 \% and 0.9\% for the filling factor, $3\degree$ and $2\degree$ for the field inclination, and $-40\degree$ and $-3\degree$ for the field azimuth, respectively. The corresponding rms values are:  30 G and 10 G, 2.2 \% and 1.7\%, $25\degree$ and $12\degree$, and $45\degree$ and $40\degree$. The larger values for the azimuth errors are the result of the very low linear polarization signal produced by these very weak fields. Moreover, the shapes of the various distributions are fully compatible to those in Fig.\ \ref{fig:Fig_B1500}, calculated from Eq.\ (\ref{eq:variancex}). Obviously, since we have here used a broader range of values for the various model parameters, the scatter of the different panels tends to be slightly larger than that for the theoretical predictions. 

\acknowledgements 

This work has been partially funded by the Spanish Ministerio de Educaci\'on y Ciencia, through Projects No. ESP2006-13030-C06-02 and PCI2006-A7-0624, and Junta de Andaluc\'{\i}a, through Project P07-TEP-2687,  including a percentage from European FEDER funds.

\bibliographystyle{aa}

\clearpage







\end{document}